\documentclass[pdflatex,sn-nature, referee]{sn-jnl}

\usepackage{graphicx}%
\usepackage{multirow}%
\usepackage{amsmath,amssymb,amsfonts}%
\usepackage{amsthm}%
\usepackage{mathrsfs}%
\usepackage[title]{appendix}%
\usepackage{xcolor}%
\usepackage{textcomp}%
\usepackage{manyfoot}%
\usepackage{booktabs}%
\usepackage{newunicodechar}
\newunicodechar{⁻}{\textsuperscript{-}}
\usepackage{algorithm}%
\usepackage{algorithmicx}%
\usepackage{algpseudocode}%
\usepackage{listings}%
\usepackage{threeparttable}
\usepackage{tcolorbox}
\usepackage{makecell}
\usepackage{pdfpages}

\usepackage[utf8]{inputenc} 
\usepackage[T1]{fontenc}    
\usepackage{hyperref}       
\usepackage{url}            
\usepackage{booktabs}       
\usepackage{amsfonts}       
\usepackage{nicefrac}       
\usepackage{microtype}      
\usepackage{lipsum}		
\usepackage{graphicx}
\usepackage{natbib}
\usepackage{doi}
\usepackage{soul}
\usepackage{tcolorbox}
\usepackage{multirow}
\usepackage{booktabs}
\usepackage{makecell}
\usepackage{array}
\usepackage{amsmath}
\usepackage{eurosym}
\usepackage{rotating}

\usepackage[utf8]{inputenc}
\usepackage{textcomp}
\DeclareUnicodeCharacter{2212}{\textminus}

\usepackage{caption}

\theoremstyle{thmstyleone}%
\theoremstyle{thmstyletwo}%

\theoremstyle{thmstylethree}%

\begin{document}

\title[Article Title]{Generative Molecular Design with Steerable and Granular Synthesizability Control}


\author*[1,2]{\fnm{Jeff} \sur{Guo}}\email{jeff.guo@outlook.com}
\equalcont{These authors contributed equally to this work.}

\author[1,2,3]{\fnm{Víctor} \sur{Sabanza-Gil}}
\equalcont{These authors contributed equally to this work.}

\author[4]{\fnm{Olha} \sur{Semenenko}}
\author[4,8]{\fnm{Oleksii} \sur{Hrabovskyi}}
\author[4,6]{\fnm{Mykola} \sur{Protopopov}}
\author[4,6]{\fnm{Anna} \sur{Kapeliukha}}
\author[4]{\fnm{Oleksandr} \sur{Mosia}}
\author[5]{\fnm{Sofiia} \sur{Hatych}}
\author[5]{\fnm{Diana} \sur{Alieksieieva}}

\author[2,3]{\fnm{Tom} \sur{Nelis}}
\author[3]{\fnm{Patrick} \sur{Molliet}}
\author[2,9]{\fnm{Helena} \sur{Solé-Àvila}}
\author[2,9]{\fnm{Valentas} \sur{Olikauskas}}
\author[2,9]{\fnm{Nina} \sur{Aregger}}
\author[10]{\fnm{Irina} \sur{Morozova}}
\author[10]{\fnm{Joseph} \sur{Schmidt}}
\author[1]{\fnm{Zlatko} \sur{Jončev}}

\author[4]{\fnm{Olga} \sur{Tarkhanova}}
\author[5]{\fnm{Petro} \sur{Borysko}}
\author[9]{\fnm{Jerome} \sur{Waser}}
\author[10]{\fnm{Bruno} \sur{Correia}}
\author[2,3]{\fnm{Jeremy} \sur{Luterbacher}}
\author*[1,2]{\fnm{Philippe} \sur{Schwaller}}\email{philippe.schwaller@epfl.ch}


\affil[1]{\orgdiv{Laboratory of Artificial Chemical Intelligence (LIAC)}, \orgname{EPFL}, \orgaddress{\city{Lausanne} \postcode{1012}, \country{Switzerland}}}
\affil[2]{\orgdiv{NCCR Catalysis}, \orgname{EPFL}, \orgaddress{\city{Lausanne} \postcode{1012}, \country{Switzerland}}}
\affil[3]{\orgdiv{Laboratory of Sustainable and Catalytic Processing (LPDC)}, \orgname{EPFL}, \orgaddress{\city{Lausanne} \postcode{1012}, \country{Switzerland}}}

\affil[4]{\orgname{CHEMSPACE LLC}, \orgaddress{\city{Kyiv}, \postcode{02094}, \country{Ukraine}}}

\affil[5]{\orgname{Enamine Ltd.}, \orgaddress{\city{Kyiv}, \postcode{02094}, \country{Ukraine}}}

\affil[6]{\orgname{Taras Shevchenko National University of Kyiv}, \orgaddress{\city{Kyiv}, \postcode{01033}, \country{Ukraine}}}

\affil[7]{\orgdiv{V. P. Kukhar Institute of Bioorganic Chemistry and Petrochemistry}, \orgname{National Academy of Sciences of Ukraine}, \orgaddress{\city{Kyiv}, \postcode{02094}, \country{Ukraine}}}

\affil[8]{\orgdiv{Palladin Institute of Biochemistry}, \orgname{National Academy of Sciences of Ukraine}, \orgaddress{ \city{Kyiv}, \postcode{01054}, \country{Ukraine}}}

\affil[9]{\orgdiv{Laboratory of Catalysis and Organic Synthesis (LCSO)}, \orgname{EPFL}, \orgaddress{\city{Lausanne} \postcode{1012}, \country{Switzerland}}}

\affil[10]{\orgdiv{Laboratory of Protein Design and Immunoengineering (LPDI)}, \orgname{EPFL}, \orgaddress{\city{Lausanne} \postcode{1012}, \country{Switzerland}}}


\abstract{Designing molecules that are both property-optimal and readily synthesizable is a central challenge in drug discovery. Existing works that \textit{do} consider synthesizability can jointly output predicted synthesis routes for generated molecules. However, there has been minimal attention in addressing the \textit{ease} of synthesis and with flexibility to incorporate desired reaction constraints. On the other hand, virtual screening searches for commercially available compounds, but imposes challenges when scaling to ultra-large (billion-size and beyond) chemical spaces. Here, we propose a generative design framework that unifies synthesis-constrained molecular design and ultra-large-scale virtual screening through \textit{steerable and granular} synthesizability control. Generated molecules satisfy arbitrary multi-parameter optimization objectives with predicted synthesis routes satisfying mix-and-match constraints: including or avoiding certain reactions, incorporating specific building blocks, and minimizing synthesis route length. In an end-to-end in-house campaign targeting BRD4, we designed molecules synthesizable with specific selected reactions and building blocks, synthesized all six selected compounds, and identified two micromolar binders. We further demonstrate that reaction control enables efficient navigation of ultra-large make-on-demand chemical spaces to identify property-optimal candidates. By applying our framework to Chemspace's Freedom 4.0 make-on-demand space (142 billion molecules), we generated $\sim$320k molecules (0.00023\% of the library) on a \textit{single} consumer-grade GPU (with only 8 GB GPU memory) and identified a micromolar Wee1 binder amongst 60 synthesized candidates. The single unified framework thus enables generating novel synthesizable molecules and retrieving catalogue-ready candidates, offering a flexible solution to mitigating the synthesizability bottleneck.}

\keywords{molecular generative design, synthesizability, ultra-large virtual screening, make-on-demand molecular libraries}

\maketitle

\section*{Main}
\label{intro}

Designing and validating molecules are the two fundamental components of molecular discovery. A key development in the last decade to augment domain experts' decision-making is enhanced capabilities of \textit{in silico} molecular design, which has been enabled by increasingly accessible compute. In the pharmaceutical industry, every prioritized molecule for experimental validation now undergoes significantly more \textit{in silico} assessment \cite{increasing-compute-in-pharma}, in light of improved predictive accuracy for properties of interest. A ubiquitous example is the prediction of binding affinity using molecular dynamics simulations over molecular docking \cite{fep-accuracy, vytautas-al-abfe, olexandr-al-fep}. This is a natural progression as experimental validation remains the major bottleneck, so greater confidence in the designed compounds can mitigate risk \cite{blakemore2018organic}. 

The biggest bottleneck for experimental validation is synthesis, typically making up most of the time in the design-make-test-analyze cycle \cite{dmta-synthesis-time}. Historically, screening physical molecular libraries such as an internal catalogue could side-step synthesizability challenges, but at the expense of a relatively constrained chemical space \cite{hts}. More recently, virtual make-on-demand libraries enumerated using robust reactions have become an attractive avenue to mitigate synthesizability challenges, particularly in early-phase hit finding. These compounds can be directly purchased from chemical vendors, which often report an 80\% synthesis success rate \cite{enamine-real}. Accordingly, many approaches cast the molecular design problem as identifying property-optimal compounds in these predefined libraries, under minimal compute. These virtual screening (VS) approaches continue to achieve widespread success \cite{stoichet-1, v-synthes} and are expected to improve with more compute and larger libraries \cite{Lyu2023-virtual-library-expansion}. However, while compute is becoming cheaper, these ultra-large-scale libraries are growing exponentially in size, reaching trillion-scale as of late 2025 \cite{emolecules-explore}. Consequently, VS endeavors can be prohibitively costly, and doing so at a reasonable cost remains a considerable challenge \cite{bigbang_chemical_libraries}.

Molecular design using machine learning (ML) has emerged as an alternative to avoid the computational expense of direct virtual screening. By implicitly learning molecular distributions, these algorithms can generate molecules efficiently without the need of exhaustively evaluating all the candidates \cite{generative-design-review}. In 2024, there were more published case studies (in academia and industry) achieving experimental validation than in the cumulative years before \cite{generative-design-review, industry-clinical-candidates}, showing the rapid increase in adoption of both methods. While promising, the synthesizability of machine-generated molecules remains a limitation \cite{gao2020synthesizability, fake-it-until-you-make-it, elephant-synthesizability}. Without the constraint of curated reactions, generated molecules can pose significant synthesis challenges that, while not necessarily impossible to synthesize, can make it practically disadvantageous over existing VS methods. Accordingly, recent ML-based methods have tackled this synthesizability bottleneck by imposing that generated molecules are predicted to be synthesizable using a predefined set of reaction rules \cite{synflownet, synformer, synthemol}. However, these approaches can be inflexible to on-the-fly changes to the synthesizability definition, e.g., such as enforcing only user-defined reactions or incorporating specific building blocks. Consequently, there remain barriers to practical adoption into laboratory workflows.

\begin{figure}[t]
    \centering
    \includegraphics[width=1.0\linewidth]{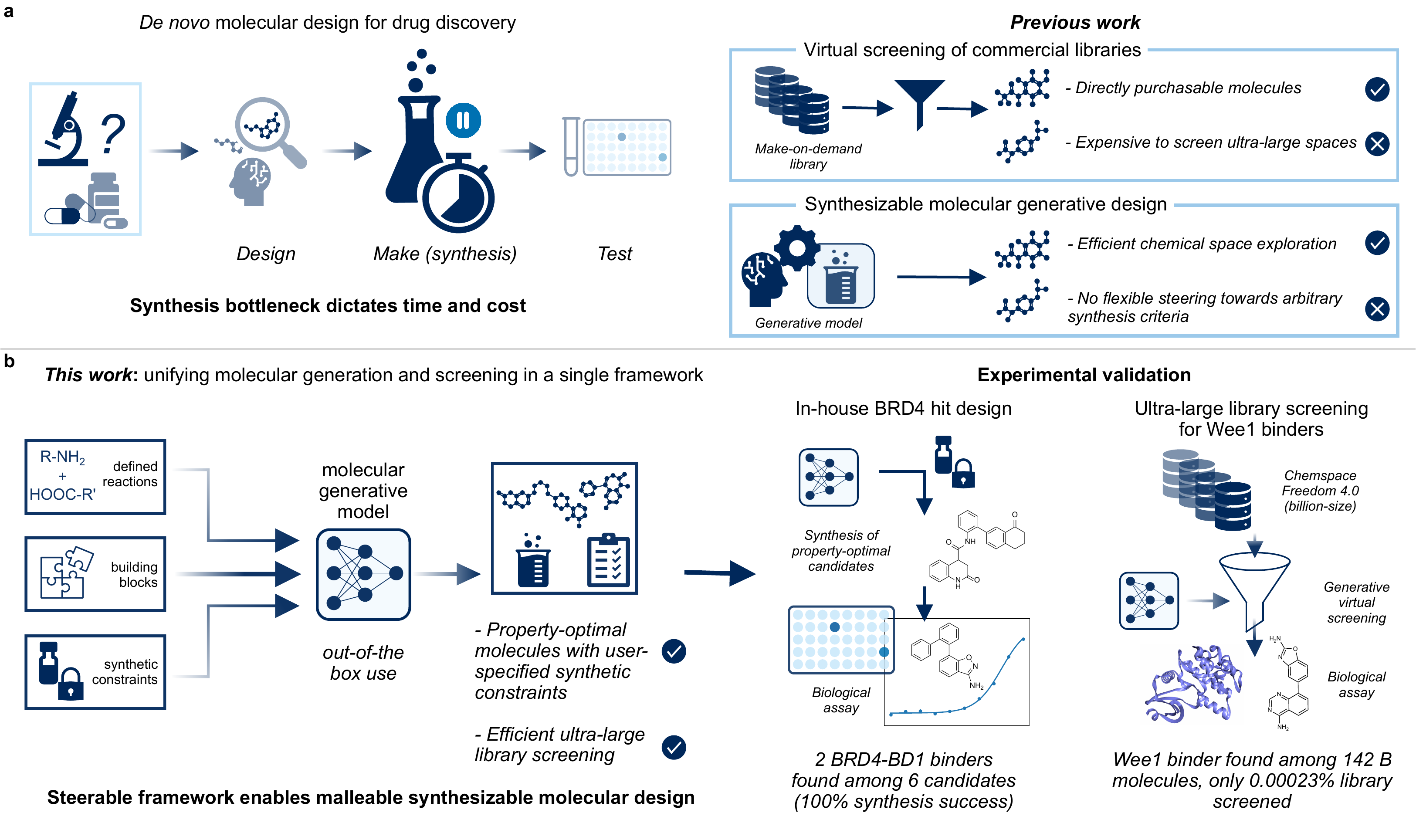}
    \caption{\textbf{Generative molecular design with synthesis constraints.}
    a) Molecular design in drug discovery is fundamentally limited by synthesis. Previous methods are less flexible to be adapted to satisfy diverse synthetic constraints. b) Our proposed framework. All generated molecules have a predicted synthesis route which enables imposing diverse constraints including enforcing specific building blocks, enforcing specific reactions, avoiding specific reactions, and minimizing the number of synthesis steps. Molecular generation can mix-and-match these constraints which enables steerable and granular design. We prospectively validated our framework with an in-house end-to-end design, synthesis, and experimental validation, yielding to $\mu M$ binders to BRD4-BD1. We adapted the same framework to perform generative virtual screening on an ultra-large make-on-demand library (Freedom 4.0, 142B molecules), finding a $\mu M$ binder to Wee1 with only $\sim$320k molecules generated (0.00023\% of the library).}
    \label{fig:intro}
\end{figure}

In this work, we introduce an open-source molecular generative framework that enables steerable and granular control over the synthesizability of generated molecules, unifying synthesizable molecular design and ultra-large virtual screening (Figure \ref{fig:intro}). We validate the practical utility of the framework in two different use cases. First, we use our model to design small molecule binders synthesizable using specific selected reactions and building blocks. Performing end-to-end design to experimental validation in-house, we generated, synthesized, and validated two hit molecules with micromolar ($\mu M$) range activity (most potent 32 $\mu M$) against the BRD4-BD1 protein. Most importantly, all the six tested molecules were synthesized with minimal changes to the predicted synthesis routes, showing direct actionability of generated candidates. Second, we adapt our framework with no computational overhead to perform retrieval of property-optimal molecules in an ultra-large-scale make-on-demand library (Chemspace's Freedom 4.0, 142 billion molecules) to search for Wee1 binders. After generating only $\sim$320k molecules (0.00023\% of the library) on a single consumer-grade GPU, 60 candidate molecules included in Freedom 4.0 were successfully synthesized and tested, resulting in a 11 $\mu M$ hit. Overall, our framework can accelerate the design-make-test-analyze cycle by unifying synthesizable generative design and ultra-large scale VS, while running on hardware available to any laboratory.

\section*{Results}
\label{results}

\subsection*{Framework development: Enforcing reaction constraints in the generative process}

We extend the Saturn \cite{saturn} generative framework by integrating retrosynthesis models (using Syntheseus \cite{syntheseus}) and reaction labelling (using Rxn-INSIGHT \cite{rxn-insight}) into the generative process (see the Methods section for details). Accordingly, generated molecules deemed synthesizable by the retrosynthesis model have an associated predicted synthesis route annotated with reaction class and names. Given arbitrary reaction constraints, e.g., generate molecules synthesizable by only amide coupling reactions, we can return a reward signal based on alignment to the imposed constraints, thus enabling steering the generative process (Figure \ref{fig:dev_results}a). We formulate a design objective to generate molecules with optimized QuickVina2-GPU \cite{autodockvina, quickvina2, quickvina2-gpu-2.1} docking scores against ClpP \cite{7uvu}, quantitative estimate of drug-likeness (QED \cite{qed}) values, constraining the number of hydrogen-bond donors (HBD), and under various reaction constraints across common medicinal chemistry transformations \cite{common-med-chem-reactions} (see the Methods section for a detailed description of the objective function). Extended Figure \ref{extended:dev_results}a shows example property-optimal generated molecules with predicted synthesis routes specifically containing Mitsunobu, Heck, and Wittig reactions. Figure \ref{extended:dev_results}b imposes the more stringent constraint that all reactions must be amide or Suzuki couplings (see Supplementary Information S6.1 for quantitative metrics). Moreover, one can incentivize generated molecules to possess shorter synthesis routes, while continuing to enforce reaction constraints (see Supplementary Information S6.3). We show that this provides practical control over potentially easier synthesis and multi-parameter optimization. In particular, there is generally a Pareto trade-off, such that molecules that can be synthesized in fewer reaction steps can have diminished property profiles. 

Going further, we extend our method to additionally satisfy constrained synthesizability \cite{tango}, where all generated molecules have predicted synthesis routes that incorporate a specific building block. Drawing inspiration from Wołos et al. \cite{waste-to-drugs}, we demonstrate industrial byproduct valorization by molecular repurposing. We generate molecules with optimized Gnina \cite{gnina} docking scores against COX-2 \cite{1cx2_target}, QED values, and under various reaction constraints (see the Methods section and Supplementary Information S7 for details on the design objective). Figure \ref{fig:dev_results}b demonstrates the repurposing of industrial byproducts from the US and EU using building blocks sourceable from Enamine US and EU stocks, respectively. By constraining to the geographic region, we show the flexibility of our approach to leverage Enamine's intra-region fast delivery time of one to two days. We highlight specific results: (1) In the US case study, all routes contain an amide reaction and build the final molecule in two steps, incorporating the selected US industrial byproduct (m-cresol, 4-hydroxyphenylacetic acid, or phenol). (2) In the EU case study, all the reactions in the syntheses are either amide or Suzuki couplings and build the final molecule in two steps, incorporating the selected EU industrial byproduct (acetic acid, 4-oxopentanoic acid, or glycine). (3) Some generated molecules incorporate a high percentage of byproduct-derived carbon atoms, which is a desirable metric when targeting bio-based products \cite{Anastas2010-12-principles}. A concrete example is the molecules derived from 4-hydroxyphenylacetic acid in the US case study which contain between 40-44\% byproduct-derived carbon atoms. It is important to note that the route quality is directly limited by the retrosynthesis model's capabilities, e.g., reaction selectivity issues. For instance, in the routes starting from glycine, selectivity issues may arise in the amide formation step due to unprotected carboxylic acid groups, and the synthesis may require adaptations. This limitation can be mitigated as retrosynthesis models improve, without changing the generative framework. Overall, these experiments show how green chemistry constraints, i.e., valorizing byproducts and minimizing the number of reaction steps \cite{Anastas2010-12-principles}, can be incorporated into \textit{de novo} design of molecules that satisfy arbitrary design objectives.

\begin{figure}[t]
    \centering
    \includegraphics[width=0.95\linewidth]{figures/Figure2.pdf}
    \caption{\textbf{Synthesis-constrained molecular generative model workflow and application}. a) A generalist molecular generative model is pretrained on 88M SMILES from PubChem \cite{pubchem}. During optimization, sampled molecules are scored against computational oracles (\textit{in silico} property predictors) defining the multi-parameter objective. One oracle, Syntheseus \cite{syntheseus}, proposes synthesis routes; if found, reactions are annotated via Rxn-INSIGHT \cite{rxn-insight} and checked against user-imposed reaction constraints (requiring, restricting, or avoiding specific reaction types). Satisfying these yields a synthesis reward of 1, aggregated with other oracle scores (e.g., RDKit objectives) to update the model via reinforcement learning (RL \cite{olivecrona-reinvent}). The loop repeats until the oracle budget is exhausted, producing property-optimal molecules with synthesis routes meeting user-specified constraints. b) \textit{In silico} industrial byproduct valorization experiment. Molecules with optimized COX-2 docking are generated together with their synthesis routes. Each route includes an enforced industrial byproduct molecule (either from the USA or the EU), in addition to other synthetic constraints.}
    \label{fig:dev_results}
\end{figure}
\clearpage

\subsection*{Prospective end-to-end in-house design, synthesis, and experimental validation of novel BRD4 binders}

\begin{figure}[!ht]
    \centering
    \includegraphics[width=1.0\linewidth]{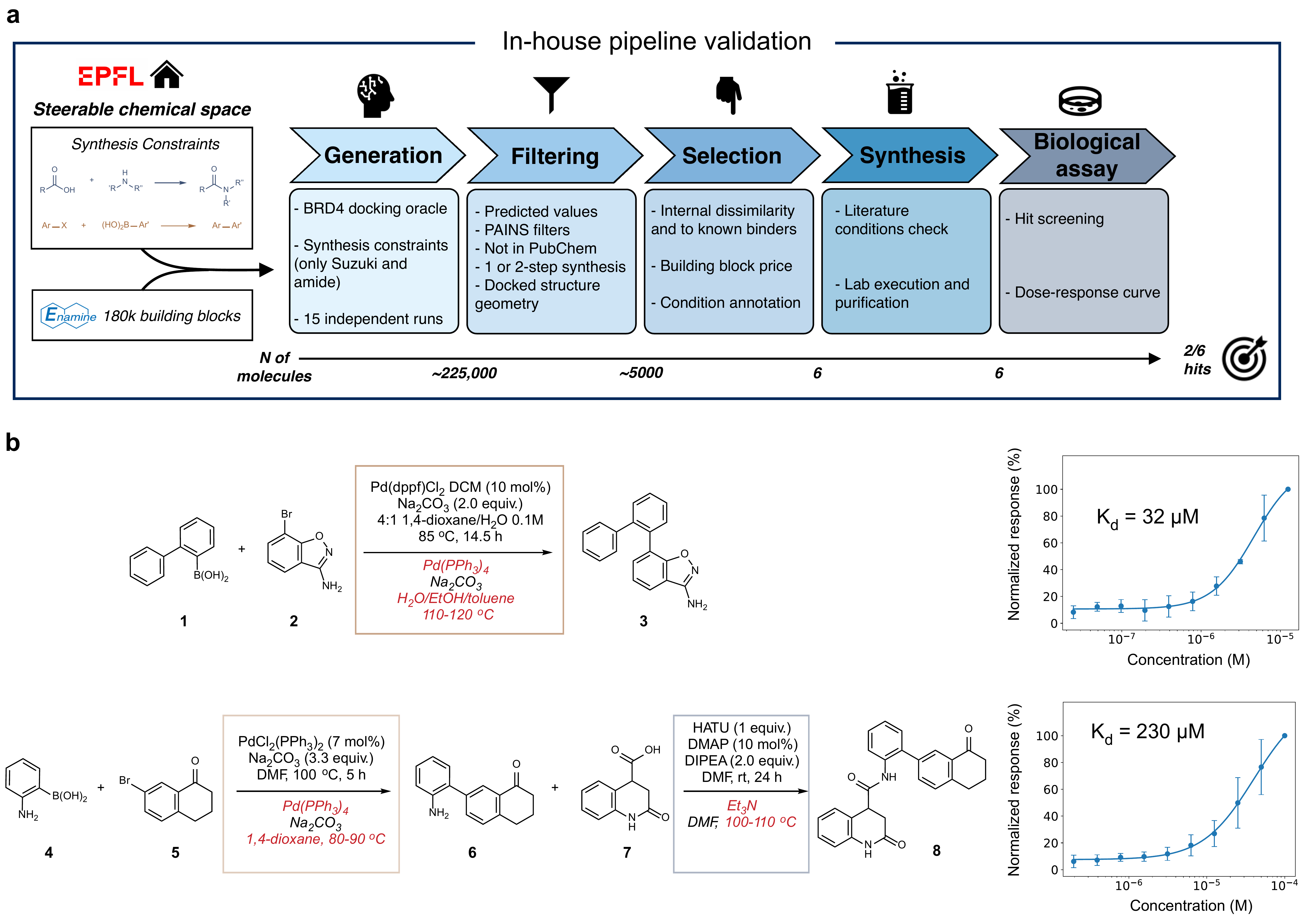}
    \caption{\textbf{BRD4-BD1 binder design experimental validation.} a) Generative pipeline for in-house experimental validation. Only amide and Suzuki reactions were allowed in the generated synthesis routes using Enamine EU building blocks. Generated molecules were optimized against a validated docking oracle to predict binding affinity (see Supplementary Information S8) and QED. Each generative experiment had a budget of 15,000 oracle calls across 5 independent seeds and with 3 different reaction constraints (allow only amide, only Suzuki, or both). This resulted in around 225,000 generated molecules in total which were filtered based on binding affinity (less than -8 kcal/mol according to the docking oracle), PAINS filters \cite{Baell2010-pains}, not in PubChem, short synthesis routes (2 or less steps), and docked poses. Final molecules were further filtered by price and dissimilarity to known binders and between selected molecules. Reaction conditions were predicted using QUARC \cite{Sun2025-quarc}, and compared to literature data to select suitable conditions. Finally, six molecules are synthesized and tested against BRD4-BD1 in a binding assay. b) Synthesis and dose-response curve of hits. Molecules 1 and 2 were synthesized according to the generated routes, and tested, resulting in $\mu M$ hits (32 $\mu M$ and 230 $\mu M$, respectively). Molecule 2 was synthesized in 2 steps, and the predicted sequence was inverted based on expert chemist assessment. Conditions predicted by the condition prediction model are shown below each reaction arrow, and marked in red if the actual conditions used in the synthesis were different. The actual conditions used in the experiments are shown above the arrow.}
    \label{fig:brd4}
\end{figure}

We experimentally validated our framework through an integrated wet-lab pipeline targeting the design of protein binders for the Bromodomain and Extra-Terminal (BET) family \cite{Baud2014-jq1, Divakaran2023-brd4-selectivity}. Specifically, we focused on the first bromodomain of BRD4 (BRD4-BD1), a well-characterized epigenetic target implicated in cancer cell proliferation and acute inflammatory responses \cite{Liu2017-brd4-review, Jiang2019-BRD4binders}. All computational, synthesis, and binding assay experiments were run in-house at the same institution. This application demonstrates the utility of synthesis-constrained generative modeling in streamlining the transition from computational design to experimental validation (Figure \ref{fig:brd4}).

The generative pipeline (Figure \ref{fig:brd4}a) utilized a validated docking oracle to estimate binding affinities to BRD4-BD1 \cite{Jiang2019-BRD4binders}. It achieves an enrichment factor of 11 with the top 0.5\% of docked molecules in a set of active and decoys, indicating the oracle's ability to discriminate potential BRD4-BD1 binders (from Jiang et al. \cite{Jiang2019-BRD4binders}) and non-binders (see Supplementary Information S8.1 for further information about the docking validation). To ensure rapid experimental turnaround, we restricted the generative chemical space to molecules reachable via amide and Suzuki-Miyaura couplings due to their high robustness and prevalence in medicinal chemistry \cite{common-med-chem-reactions}. Building blocks were sourced exclusively from the Enamine EU stock (September 2025, 180k blocks) for 1-2 day delivery \cite{enamine_bb_catalog_2025} within the EU. In addition to binding affinity, the multi-objective optimization included QED \cite{qed} to maintain favorable physicochemical properties (see Supplementary Information S8.2 for a complete description of the MPO target).

We set up three different generative experiments varying the allowed chemical reactions. In the first one, we allowed both amide and Suzuki reactions, in the second only amide, and in the third only Suzuki. We ran five independent seeds of 15,000 generated molecules for each experiment, generating an ensemble of $\sim$225,000 molecules in 15 independent runs. We implemented a post-hoc filtering cascade to select molecules for experimental validation. First, we retained only molecules with predicted docking scores $< -8$ kcal/mol (according to our validated protocol) that are not present in PubChem \cite{pubchem}. Next, we applied Pan-Assay Interference Compounds (PAINS) filters \cite{Baell2010-pains} to remove likely false positives, and enforced a conserved hydrogen-bond with the key ASN140 residue as a structural integrity criterion \cite{Divakaran2023-brd4-selectivity}. We further narrowed the set by requiring synthetic accessibility of $\le 2$ steps with an estimated cost of $<$ \euro300 per route. Finally, to ensure chemical diversity, we imposed a strict Tanimoto similarity threshold of $< 0.5$ both relative to known BRD4 ligands \cite{Liu2017-brd4-review} and amongst the selected candidates themselves. For additional details about the computational funnel, see Supplementary Information S8.3.

Six candidates were prioritized for synthesis and biological evaluation (Extended Figure \ref{extended:synthesis}). As Syntheseus only provides a predicted synthesis route, we annotated reaction conditions using a condition prediction model \cite{Sun2025-quarc} and subsequently refined through expert chemist review. Notably, all six selected molecules were successfully synthesized following the constrained 1- and 2-step routes (see Supplementary Information S8.4 for compound synthesis and characterization details). Although one-step reactions could in principle be treated as enumeration, we estimated the total number of possible combinations for the amide and Suzuki coupling templates at $\sim$3.2 billion and $\sim$300 million candidates, respectively. Consequently, the 1- (and by extension the 2-step) synthetic spaces are too vast for exhaustive enumeration to be computationally easily tractable. Competitive binding assays against BRD4-BD1 confirmed two hits, yielding a 33\% hit rate (Figure \ref{fig:brd4}b). The binding affinities for molecules 1 and 2 are $32\mu M$ and $230 \mu M$, respectively (see Supplementary Information S8.5 for a description of the primary screening and dose response experiments). Micromolar binders represent validated starting points for lead optimization, showing a fast translation between molecule generation and experimental testing. These results demonstrate that incorporating synthetic constraints directly into the generative process can yield chemically novel, synthetically accessible, and biologically active scaffolds, significantly reducing the experimental iteration cycle time, while using fully open-source software.

\subsection*{Generative virtual screening of an ultra-large make-on-demand commercial library finds hits against Wee1}

Ultra-large make-on-demand libraries have reached trillion-compound scale \cite{ultra-large-databases}. Identifying property-optimal compounds in these libraries is a common and attractive endeavour, especially in early-phase hit finding, because it allows one to mitigate synthesizability challenges. By leveraging robust combinatorial chemistry, make-on-demand libraries commonly report 80\% synthesis success rate with a delivery time of under a month \cite{enamine-real}. However, it is infeasible to exhaustively screen these libraries. A recent study reported using 5.6 million CPUs to dock a 69-billion molecular library \cite{virtualflow2}. Another study suggested that the average cost of docking 1 billion molecules exhaustively is around \$12,000 \cite{thompson-sampling-vs}. Correspondingly, the field has developed approaches to minimally screen these libraries, while retrieving property-optimal compounds: (1) Synthon-based approaches which consider the minimal fragments used to combinatorially enumerate the library. By focusing only on promising synthons (based on the objective function), the space can be largely pruned. Approaches such as V-SYNTHES \cite{v-synthes} and Thompson sampling \cite{thompson-sampling-vs} screen only 2 million compounds out of 11 billion and 3.35 million compounds out of 335 million (reported 1\%). (2) Generative approaches that generate specifically in make-on-demand space. Approaches such as SyntheMol \cite{synthemol} and NeuralGenThesis (NGT) \cite{atomwise-ngt} define the generative process to use the same make-on-demand library's building blocks and reactions or learns a latent representation of the space, respectively. Other generative approaches like SynFormer \cite{synformer} and GFlowNets \cite{rgfn, synflownet, rxnflow} may use the same building blocks and reactions, but are more lenient with the number of synthesis steps, thus generating molecules that may not be strictly commercially available.

\begin{figure}[!ht]
    \centering
    \includegraphics[width=0.90\linewidth]{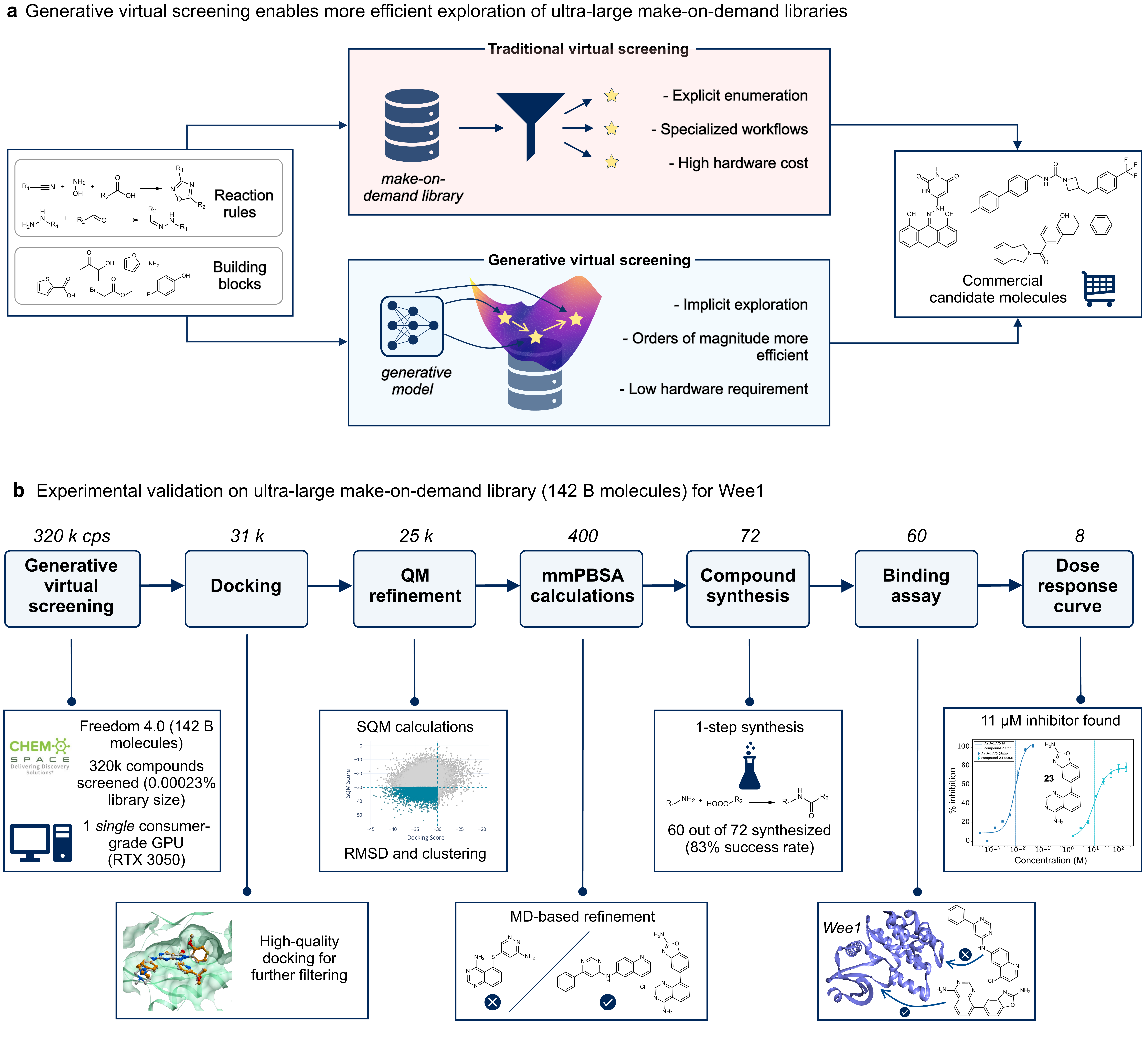}
    \caption{\textbf{Generative virtual screening (GVS) for Wee1 binder design.} a) Comparison between traditional virtual screening and GVS. By imposing similar reaction constraints to how make-on-demand libraries are enumerated, the generative process can be steered to generate exact matches in the same space. b) Prospective validation of GVS in Freedom 4.0 (142B molecules) make-on-demand space. On a single consumer-grade GPU, we generated 320,045 molecules which amounts to 0.00023\% of the library size. Depending on the specific imposed reaction constraints, over 95\% of generated molecules can be exact matched in Freedom 4.0. These exact matches were triaged through more rigorous docking, semi-empirical calculations, and molecular dynamics simulations. The most promising subset were synthesized and tested, resulting a 11 $\mu M$ hit.}
    \label{fig:gvs}
\end{figure}

Here, we show that our approach can be used out-of-the-box to efficiently retrieve property-optimal compounds in ultra-large make-on-demand libraries. Since these libraries are combinatorially enumerated with a fixed set of building blocks and reactions, we hypothesized that mimicking similar constraints in the generative process will allow retrieval (Figure \ref{fig:gvs}a). In Extended Data Table \ref{tab:generative-virtual-screening}, we show that changing the enforced reaction changes the exact match rate, i.e., how many generated molecules are found in the make-on-demand library (Chemspace's Freedom 4.0). When enforcing only amide reactions, we observe that 97\% of generated molecules satisfying this reaction constraint are matched in Freedom 4.0. See Supplementary Information S9 for additional results and implications of enforcing different reactions. Next, we ran two \textit{in silico} benchmarks against existing approaches: (1) Use our same pretrained model on PubChem to compare against SynFormer and GFlownet approaches. Supplementary Information S11 shows that our approach generates more property-optimal compounds (docking, QED, HBD, and predicted to be synthesizable) under both oracle evaluation and compute time budgets. (2) Removing the confounding variable of different pre-training data, we pretrained Saturn and SynFormer specifically on Chemspace's Freedom 4.0 space and adapted SyntheMol to generate in the same space. Supplementary Information S11 shows that our approach also generates more property-optimal compounds (docking, QED, HBD, and predicted to be synthesizable). After validating our approach's feasibility and optimization efficiency, we demonstrate the generative virtual screening (GVS) of Freedom 4.0 (142 billion space). With no extra computational overhead, we used the same pretrained model on PubChem to retrieve property-optimal compounds by docking only 15,000 compounds, representing 0.000011\% of the spaces (see Extended Data Table \ref{tab:generative-virtual-screening}).


Subsequently, we prospectively applied our framework to identify Wee1 inhibitors in Chemspace's Freedom 4.0 space. Wee1 is a serine/threonine and tyrosine kinase that regulates the G2/M cell cycle checkpoint through CDK1 phosphorylation, implicated as a promising oncology drug target because its inhibition forces cancer cells into premature mitosis with unrepaired DNA damage, leading to mitotic catastrophe and cell death \cite{Bukhari2022-wee1, GhelliLusernadiRor2020-wee1}. Figure \ref{fig:gvs}b shows the computational pipeline where we first task Saturn to generate docking-optimal compounds in Freedom 4.0 (see the Methods section for full details). Using a single consumer-grade GPU (NVIDIA RTX 3050, 8 GB memory), we generated and docked 320,045 compounds (0.00023\% of the Freedom 4.0 space). This highlights the model's minimal compute requirements for navigating ultra-large chemical spaces. Next, we filtered the generated molecules with three different design objectives related to the docking protocol: (1) Favorable docking score only. (2) Docking score and hydrogen-bond interactions to C379, E377, and N376. (3) Docking score, hydrogen-bond interactions, and pose. The resulting molecules were refined using a computational funnel consisting of further docking, pose refinement, semi-empirical quantum mechanical (SQM) computations\cite{ez2016-cuby} and clustering, and molecular mechanics Poisson-Boltzmann surface area (mmPBSA) simulations \cite{ValdsTresanco2021_gmxMMPBSA} (see Supplementary Information S10 for full details). The funnel prioritized 72 candidates for synthesis. 60 out of 72 molecules were successfully synthesized according to the predefined protocols (agreeing with the average 80\% success rate for the molecules in the library \cite{Kapeliukha2025-freedom}). The 60 candidates were evaluated in a primary screening assay, yielding 3 hits according to the selection criteria (more than 30\% activity). The three hits, together with the 5 molecules closest to the threshold were subjected to dose-response experiments. Compound \textbf{23} was identified as a Wee1 binder, with an IC$_{50}$=11 $\mu M$. The docked pose and structure of this molecule are shown in Supplementary Figure S12, displaying three potential hydrogen-bonds through the aromatic nitrogen, amine, and amine group with E346, C379, and Y378, respectively. The docking pose suggests that the molecule was successfully localized as a strong interacting candidate in the Wee1 pocket. Overall, these result validate the use of our framework for fast identification of protein binders in ultra-large commercial libraries.

\section*{Discussion}
Synthesizability is the major bottleneck in the design-make-test-analyze cycle. Promoting synthesizability in generated molecules is necessary to make \textit{in silico} design actionable. On the other hand, the challenge of synthesizability can be mitigated by performing property-optimal molecule retrieval in ultra-large-scale make-on-demand libraries. However, this approach can incur a prohibitive computational cost. In this work, we proposed a small molecule generative framework that unifies synthesis-constrained molecular design and ultra-large-scale virtual screening through \textit{steerable and granular} synthesizability control. Previous work assessed synthesizability using the proxy of whether a retrosynthesis model successfully predicts a synthesis route, and found that only about 30\% of generated molecules satisfy this constraint \cite{gao2020synthesizability}. By integrating retrosynthesis models and reaction labelling tools \textit{directly} in the generative process, we ensure all property-optimal generated molecules \textit{must} contain a predicted synthesis route. Using this framework, we demonstrated the capability to lower the synthesizability barrier through two prospective experimental case studies: (1) In an end-to-end in-house design, synthesize, and binding validation campaign, we showed that generated molecules predicted to be synthesizable using only amide and Suzuki couplings are directly actionable. Using the predicted synthesis routes with minimal changes, we successfully synthesized six out of six compounds and identified two hit molecules (most potent 32 $\mu M$) against the BRD4-BD1 protein. Importantly, this mode of control is aligned with a practical strategy to derisk synthesis by leveraging a small number of robust reactions with diverse building blocks \cite{common-med-chem-reactions}. (2) Using a single consumer grade GPU, we performed generative virtual screening of Chemspace's Freedom 4.0 make-on-demand space comprised of 142 billion molecules to identify Wee1 binders. By generating only 320,045 molecules (0.00023\% of the library), we retrieved 72 property-optimal candidates, of which 60 were successfully synthesized, leading to an 11 $\mu M$ hit. Our framework facilitates experimental iteration and importantly, demonstrates that synthesizability control jointly enables actionable generative design and efficient retrieval in ultra-large make-on-demand libraries.

Next, we highlight limitations of our method and opportunities for future development. First, our framework integrates retrosynthesis models as a feedback signal and thus the reliability of the synthesis routes strictly depend on the model's chemical accuracy. Consequently, selectivity or reactivity problems can currently arise \cite{imperfect-reaction-templates}. Another limitation of our method is the use of a binary reward signal, i.e., the generative model receives a reward of 1 if the reaction constraints are satisfied, otherwise 0. This binary and sparse reward can be challenging to navigate and is observed when enforcing less common reactions. One approach to mitigate this is by seeding the replay buffer with a pre-enumerated set of molecules that already satisfy the reaction constraint(s). We experimented with this in Supplementary Information S9 and observed no discernible improvement, but expect that it can offer performance benefits for other objective functions. 

An interesting future avenue is to consider multiple predicted synthesis routes for each generated molecule. This may mitigate reward sparsity as there would be more chances that at least one of the routes satisfies the reaction constraint(s). Another interesting direction is to better understand the reaction constraint dynamics when navigating make-on-demand space. Given that certain sub-spaces of these libraries are enumerated from specific reactions, enforcing the same reactions in the generative process may allow retrieving from these chemical space pockets (see Supplementary Figure S6d). While we performed some experiments to answer this question in Supplementary Information S9.2, understanding the granularity in which this can be done will have practical implications on the limits of generative virtual screening.

Overall, we demonstrate that a general pretrained model can flexibly satisfy arbitrary synthesizability constraints. Our framework jointly tackles generating synthesizable molecules and efficiently retrieving property-optimal molecules in ultra-large-scale make-on-demand spaces. By prospectively validating our framework on both practical use cases, we demonstrate a meaningful step towards alleviating the synthesizability bottleneck in generative molecular design.

\section*{Methods}
\label{Methods}
\subsection*{Molecular generative framework}
We build on Saturn \cite{saturn} which is a language-based molecular generative framework using the Mamba \cite{mamba} architecture and pretrained on the PubChem \cite{pubchem} dataset in the SMILES \cite{smiles} format (see Supplementary Information S2 for pre-training details). Next, this pretrained model is tuned with reinforcement learning (RL) to achieve goal-directed generation, i.e., tailoring the generation of molecules towards user-specified property objectives. The specific RL algorithm is Augmented Memory \cite{augmented-memory}, which we previously proposed by building on REINVENT's algorithm \cite{olivecrona-reinvent, reinvent2, reinvent4}, which in turn is a variant of REINFORCE \cite{reinforce}. Notably, Saturn is an unconditional, i.e., pre-training with unlabeled data and unconstrained, i.e., generation always proceeds token-by-token with no inductive biases. We show that incentivizing general-purpose models with a sample-efficient and malleable algorithm can successfully tackle challenging molecular design objectives. See Supplementary Information S1 for details about the Mamba architecture and RL algorithm. 

In this work, we use Syntheseus \cite{syntheseus}, which is a wrapper around single-step retrosynthesis models and search algorithms, allowing mix-and-match formulations to tackle multi-step retrosynthesis. Through Syntheseus, we use the MEGAN graph-edits \cite{megan} model re-trained by the Syntheseus authors on USPTO-50k \cite{uspto, uspto-split}. The search algorithm is Retro* \cite{retro*} with a search time limit of 180 seconds and the retrosynthetic expansion only considers the top-1 reaction by probability. In practice, considering only the top-1 reaction results in fast retrosynthetic search as the expansion graph does not grow large. For the development experiments, we used the eMolecules building block stock (23,077,162) \cite{retro*}. For other experiments, the building block sets were case-specific. Syntheseus outputs a solved synthesis route if the retrosynthetic search successfully decomposes the target molecule into available precursors, as defined by the building block set. To introduce reaction constraints, we adopt a straightforward approach to label these synthesis routes with reaction class and name information. Every chemical transformation from Syntheseus is represented by a reaction SMILES. We use Rxn-INSIGHT \cite{rxn-insight} which takes as input reaction SMILES and performs 10-class classification using bond-electron matrices, with the reaction class nomenclature following \cite{10-reactions}. Next, the reaction name is retrieved by matching to a set of 527 reaction SMIRKS \cite{smirks}. In addition, other reaction classification software like NameRxn \cite{namerxn} can be also used.

Previously, by running an explicit retrosynthesis search on every generated molecule, we showed that Saturn can generate property-optimized molecules with a predicted synthesis route \cite{saturn-synth}, while also optionally enforcing the presence of user-specified building blocks \cite{tango}. Here, we extend these capabilities to also enforce or avoid user-specified reaction classes or even \textit{specific} reactions. We introduce a simple synthesizability binary reward, such that the model receives a perfect reward (1.0) if the user-specified reaction constraint is met and 0.0 otherwise. We write `constraints` to encompass the following:

\begin{enumerate}
    \item \textbf{`Enforce Reaction`}: The user-specified reactions are incorporated ($\geq$ 1) in the synthesis routes.
    \item \textbf{`Avoid Reaction`}: The user-specified reactions are \textit{not} present in the synthesis routes.
    \item \textbf{`Enforce all Reactions`}: All reactions belong to the user-specified reactions set.
\end{enumerate}

Note that one could enforce the presence of certain reactions while simultaneously avoiding others, as we show in the Results section. Moreover, the `Enforce all Reactions` constraint implies that other reactions are avoided.

Our framework can be adapted out-of-the-box for diverse use cases without the need for additional pretraining or model changes. For example, in the in-house BRD4-BD1 prospective validation case study, we changed the building block stock of the retrosynthesis model to use Enamine EU (1-2 days delivery within the EU), and enforced only the reaction types we wanted to run in the lab (in this case, amide and Suzuki couplings). For the Wee1 generative virtual screening case study, we changed the building block stock to Chemspace Freedom 4.0's specific building blocks, and enforced the reactions that were used to define the space, such as amide coupling.

\subsection*{Framework development: Design objectives and metrics}

\subsubsection*{Enforcing reaction constraints in the generative process}
For method development, we fixed a multi-parameter optimization (MPO) objective relevant to drug discovery. The components of the objective function are:

\begin{enumerate}
    \item \textbf{Docking:} Minimize QuickVina2-GPU \cite{autodockvina, quickvina2, quickvina2-gpu-2.1} docking scores against ATP-dependent Clp protease proteolytic subunit (ClpP) \cite{7uvu}.
    \item \textbf{QED:} Maximize the quantitative estimate of drug-likeness (QED) \cite{qed} score.
    \item \textbf{Hydrogen-bond Donors:} Constrain the number of hydrogen-bond donors < 4 which could improve ADME properties \cite{hbd-metric, hbd-in-drug-design}.
    \item \textbf{Synthesizability:} Various degrees of synthesizability based on predicted synthesis routes from Syntheseus \cite{syntheseus} and reaction labels from Rxn-INSIGHT \cite{rxn-insight}.
\end{enumerate}

We quantitatively assessed the compound quality based on the evaluation metrics summarized in Table \ref{tab:general-metrics}. 

\begin{table}[!ht]
\centering
\caption{Summary of general evaluation metrics. The mean and standard deviation are reported across all 5 (0--4 inclusive) seeds. For the \textbf{Molecule Set Quality (Pooled)} metrics, all molecules across the 5 replicates are pooled, de-duplicated, and statistics reported.}
\label{tab:general-metrics}
\begin{tabular}{@{}p{0.25\linewidth}p{0.70\linewidth}@{}}
\toprule
\textbf{Category} & \textbf{Metric (averaged or pooled across all 5 replicates)} \\
\midrule
\multirow{6}{=}{\textbf{Synthesizability}} 
  & \textbf{Non-synth}: retrosynthesis model does not return a synthesis route\\[3pt]
  & \textbf{Synth}: synthesis route returned but not all constraints are met \\[3pt]
  & \textbf{Synth (constraints)}: synthesis route returned and all constraints satisfied \\[3pt]
  & \textbf{Rxn Steps}: \# reaction steps amongst all Synth (constraints) \\[3pt]
  & \textbf{Enforced Blocks} (reported if applicable): \# unique enforced blocks amongst all Synth (constraints) \\[3pt]
  & \textbf{N Successful Runs}: \# seeds (out of 5) with $\geq$ 1 Synth (constraints) \\
\midrule
\multirow{6}{=}{\textbf{Molecule Set Quality (Pooled)}} 
  & \textbf{QuickVina2-GPU Docking Scores} \cite{autodockvina, quickvina2, quickvina2-gpu-2.1}: $< -10$ is considered favourable following previous work \cite{rgfn} \\[3pt]
  & \textbf{QED} \cite{qed}: Quantitative estimate of drug-likeness \\[3pt]
  & \textbf{Ligand Efficiency}: Docking score divided by \# heavy atoms \\[3pt]
  & \textbf{Bemis--Murcko Scaffolds} \cite{bemis-murcko}: \# unique Bemis--Murcko scaffolds \\[3pt]
  & \textbf{IntDiv1} \cite{moses}: Pairwise Morgan fingerprint similarity (radius=2, nBits=1024) \\[3pt]
  & \textbf{\#Circles} \cite{circles}: Sphere packing number (0.75 similarity threshold) \\
\midrule
\multirow{2}{=}{\textbf{Compute}} 
  & \textbf{Oracle Calls}: \# unique generated molecules scored by the objective function. \textit{This is fixed at 15,000}. \\[3pt]
  & \textbf{Wall Time}: Compute time on a shared NVIDIA L40S GPU cluster \\
\bottomrule
\end{tabular}
\end{table}

These metrics are used in all method development experiments and were chosen for a comprehensive coverage to answer the following questions: (1) Do generated molecules satisfy all user-specified reaction constraints? (2) Are generated molecules property-optimized? We report both \textbf{Oracle Calls} and \textbf{Wall Time} to comment on resource usage. Given a fixed oracle budget, the wall time differs across models, across reward functions, e.g., if a larger building block stock were used for retrosynthesis as the search time would vary, and across GPU types. Oracle calls may also impose additional costs such as API costs. All Tanimoto similarity calculations are computed with Morgan fingerprints with radius 2 and 1024 bits. The metrics when enforcing synthesizability using only certain reactions with and without also enforcing specific building blocks are presented in Extended Data Tables \ref{table:enforce-all-metrics} and \ref{table:enforce-all-and-blocks-metrics}, respectively. Full metrics for this section are presented in the Supplementary Information S6.1.

Finally, all wall times are reported in the Supplementary Information Table S1 and contrasts the wall time differences when using Rxn-INSIGHT compared to NameRxn for reaction labeling. See Figure \ref{fig:dev_results} and Supplementary Information S5 for example property-optimal generated molecules and their synthesis routes when enforcing varying reaction constraints.

\subsubsection*{Industrial byproducts valorization by molecular repurposing}
The design objective in this section was as follows:

\begin{itemize}
    \item \textbf{Docking}: Minimize Gnina \cite{gnina} docking scores against COX-2 \cite{1cx2_target}. We used Gnina here instead of QuickVina2-GPU due to the higher performance of the former in distinguishing actives and decoys (see Supplementary Information S7).
    \item \textbf{QED}: Maximize QED.
    \item \textbf{Synthesizability oracle}:
    \begin{itemize}
        \item US byproducts case study: Enamine US stock (Feb. 2025, 231,472 molecules) as the building block stock for the retrosynthesis model. Enforce the presence of a US industrial byproduct, presence of amide coupling, avoid protections and deprotections, and minimize the number of synthesis steps.
        \item EU byproducts case study: Enamine EU stock (Feb. 2025, 156,510 molecules) as the building block stock for the retrosynthesis model. Enforce the presence of an EU industrial byproduct and enforce that all the reactions in the synthesis routes are amide or Suzuki couplings.
    \end{itemize}
\end{itemize}

\subsection*{Prospective end-to-end in-house design, synthesis, and experimental validation of novel BRD4 binders}
To validate a docking protocol for BRD4 BD1, we used Gnina \cite{gnina} on the 6JJ3 \cite{Jiang2019-BRD4binders} crystal structure alongside 60 ligands with experimentally measured binding affinities from the same publication \cite{Jiang2019-BRD4binders}. Redocking preserved critical water molecules and the key ASN140 hydrogen-bond, achieving an RMSD of 1.18 Å. Benchmarking against ChEMBL 35 decoys—matched for molecular weight, LogP, TPSA, hydrogen-bond donors/acceptors, rotatable bonds, and formal charge—yielded an enrichment factor of 11 at the top-0.5\% and a Spearman correlation of 0.327 with experimental binding affinity, prompting the use of a reverse sigmoid reward function within the Saturn framework. Generative experiments then optimized the GNINA docking score and QED for drug-likeness across three synthetic constraint configurations: amide and Suzuki, amide only, and Suzuki only. Running five experiments per configuration with 15,000 oracle calls each generated approximately 225,000 molecules.

These molecules were filtered through a strict seven-step computational pipeline requiring: QED > 0.4 and GNINA score < -8, removal of duplicates, exclusion of PAINS-flagged compounds \cite{Baell2010-pains} and rings larger than 9 atoms, and a Tanimoto similarity < 0.5 to reference binders. The filter also enforced one- or two-step synthesis rules depending on the configuration, required docked poses with an ASN140 hydrogen-bond and an interaction with water 324, and excluded known PubChem molecules (as of October 2025) to ensure novelty. Ultimately, six distinct molecules (two per configuration) were selected based on optimal docking scores, a mutual Tanimoto similarity < 0.5, and a synthesis cost under \euro300 via the Enamine EU (September 2025) catalogue. All six were successfully synthesized, with expert chemists only swapping the step order for the two-step syntheses while maintaining the predicted reactions and final products. See Supplementary Information S8 for a detailed description of the BRD4-BD1 target and the filtering funnel.

\subsection*{Generative virtual screening of an ultra-large make-on-demand commercial library finds hits against Wee1}
\label{methods-gvs}
\subsubsection*{Synthon-based search}
Synthon-based search \cite{rdkit-synthon-search} was used as an efficient method to check for make-on-demand library membership of generated molecules. This approach requires the synthon files which define how the space's building blocks can be combinatorially enumerated following defined reactions. Accordingly, we obtained these files for Chemspace's Freedom 4.0 space. 

\subsubsection*{Generating molecules in make-on-demand space}

Virtual libraries are enumerated based on pre-selected building blocks and reactions. Conceptually, imposing the exact same reaction constraints on a generative model should enable generating directly in the same space. Accordingly, we demonstrate that enforcing similar reaction classes to the ones used to define the make-on-demand space enables exact retrieval. See Extended Data Table \ref{tab:generative-virtual-screening} for quantitative metrics across enforcing different reaction classes. See Supplementary Information S9 for additional results and a discussion on the implications of enforcing different reactions.

\subsubsection*{Generative design to search Freedom 4.0 for Wee1 binders}

Cross-referencing the Extended Data Table \ref{tab:generative-virtual-screening}, an observation we made is that using synthon-search as the reward function (1.0 if in Freedom and 0.0 if not) in place of reaction constraints (1.0 if the generated molecule is predicted to be synthesizable using specific reactions, and 0.0 if not) can generate more property-optimal molecules. Subsequently, in the prospective screening of Freedom 4.0 for Wee1 hits, we decided to use the binary synthon-search reward function for two reasons: first, while reaction constraints can steer the generative process for a targeted exploration of make-on-demand space, we did not have any specific chemical space preferences since we were interested in finding any hits. Second, we were using only a single RTX 3050 GPU with 8 GB memory to run our generative framework. Running the generative model requires around 2.5 GB GPU memory and running the reaction model (MEGAN \cite{megan} in this case) takes about 2 GB GPU memory. With 8 GB memory, this meant we could only run one generative experiment at a time. By using the synthon-search as the reward function, we could omit the reaction model and run two parallel generative experiments. Accordingly, we ran a total of 13 generative experiments (amounting to 320,045 generated molecules) with the objective function optimizing for ICM-Pro docking score and Freedom 4.0 membership. Each generative experiment took an average wall time of 63 hours. The generated molecules were filtered to have < -30 docking score and we additionally considered subsets that had < -30 radial and topological convolutional neural network score (RTCNN, a scoring model in ICM-Pro), hydrogen-bonds to the oxygen or nitrogen atom of C379, hydrogen-bond to E377, and hydrogen-bond to N376. Note that no generated molecule that was in Freedom 4.0 satisfied every single criteria: < -30 docking score, < -30 RTCNN, and all 4 hydrogen-bonds. This set of generated molecules as carried forward through a computational funnel to prioritize the final set of compounds for synthesis (see the next sub-section).

\subsubsection*{Optimization target and refinement of the generated molecules}
To identify WEE1 binders, RCSB crystal structures were prepared in ICM-Pro by adding hydrogens, optimizing His, Asn, and Gln residues, and applying local minimization. Validation via ATP pocket redocking and cross-docking identified two top models; the 8BJU-based model was further refined with water molecules and validated using 24 ChEMBL actives and 2500 DeepCoy decoys to achieve an AUC of 0.87. A virtual screening of 31,000 generated compounds was executed using ICM-Pro's VLS (thoroughness 4, BPMC sampling, rigid receptor/flexible ligand), retaining the top pose and filtering for docking scores < –20 and an hydrogen-bond with the C379 hinge. The remaining 25,000 compounds underwent SQM scoring in Cuby4 (MOPAC2016 backend), calculating the total interaction energy ($\Delta E_{int} + \Delta\Delta G_{solv}$) on ICM Pro-minimized, > 7 Å truncated complexes using the PM6-D3H4X method and the COSMO2 implicit solvent model. Applying stricter thresholds (docking < –30, SQM < –30, low docked-to-minimized RMSD) and UPGMA clustering based on minimum docking scores yielded 402 cluster centers. These complexes underwent 10 ns MD simulations in GROMACS 2023.2 (Amber ff99/GAFF2, TIP3P water, 0.15 M NaCl) involving 100,000 steepest-descent minimization steps, 500 ps NVT and 500 ps NPT equilibration (310 K, 1 atm, heavy atoms restrained at 1000 kJ·mol⁻¹·nm⁻²), and NPT production runs (2 fs step, 300 kJ·mol⁻¹·nm⁻² backbone restraints, v-rescale with 0.1 ps coupling, Parrinello–Rahman with 5 ps coupling, LINCS, PME, and a 1.0 nm potential-shift-Verlet cutoff). Following MM/PBSA protocol validation on 23 actives and 25 decoys, binding free energies were averaged using gmx\_MMPBSA (PB method, PBRadii=5, radiopt=0, fillratio=4.0, indi=1.0) across 350 frames from the final 7 ns, leading to the selection of 72 compounds for synthesis based on these energies and visual pose inspection. For a detailed description of the computational pipeline, see SI S8.

\section*{Code availability}\label{code_av}
The code is available fully open-source at \url{https://github.com/schwallergroup/saturn} (DOI: 10.5281/zenodo.20444210)

\section*{Data availability}
Experimental protocols, compound synthesis, benchmarking against other generative methods, additional details about the generative workflow are available in the Supplementary Information.

\section*{Acknowledgements}\label{code_av}
This work was created as part of NCCR Catalysis (grant number 225147), a National Centre of Competence in Research funded by the Swiss National Science Foundation. V.S.G acknowledges support from the European Union’s Horizon 2020 research and innovation program under the Marie Skłodowska-Curie grant agreement N° 945363.

\section*{Author contributions}

J.G.\ and V.S.G.: conceptualization, methodology, software, investigation, visualization, writing -- original draft. O.S., O.H., M.P., A.K., O.M., S.H., and D.A.: investigation (Wee1, computational pipeline, compound synthesis and biological characterization), validation. T.N., P.M., H.S.-A., V.O., and N.A.: investigation (BRD4-BD1 chemical synthesis and experimental validation), validation. I.M.\ and J.S.: investigation (BRD4-BD1 binding assay), validation. Z.J.: investigation, visualization. O.T., P.B., and Y.M.: supervision, resources, project administration, writing -- review \& editing. J.W.\ and B.C.: supervision, funding acquisition, resources, writing -- review \& editing. J.S.L.: supervision, funding acquisition, resources, writing -- review \& editing. P.S.: conceptualization, visualization, supervision, funding acquisition, project administration, resources, writing -- original draft. All authors reviewed and edited the manuscript.

\section*{Competing interests}
O.S., M.P., A.K., O.M., O.T., P.B. and Y.M. are full-time employees at Chemspace LLC. O.H. is a full-time employee at Enamine Ltd.

\backmatter












\bibliography{sn-bibliography}
\newpage
\section*{Extended Data}

\captionsetup[table]{name=Extended Data Table}
\captionsetup[figure]{name=Extended Data Figure}

\setcounter{table}{0}
\setcounter{figure}{0}

\begin{figure}[!ht]
    \centering
    \includegraphics[width=0.95\linewidth]{figures/Extended_enforced_rxns.pdf}
    \caption{\textbf{Examples of enforced synthetic conditions}. a) Molecules with enforced presence of Mitsunobu, Heck and Witting reactions. b) Molecules with presence of amide or Suzuki reactions and a given building block.}
    \label{extended:dev_results}
\end{figure}

\begin{figure}[!ht]
    \centering
    \includegraphics[width=0.95\linewidth]{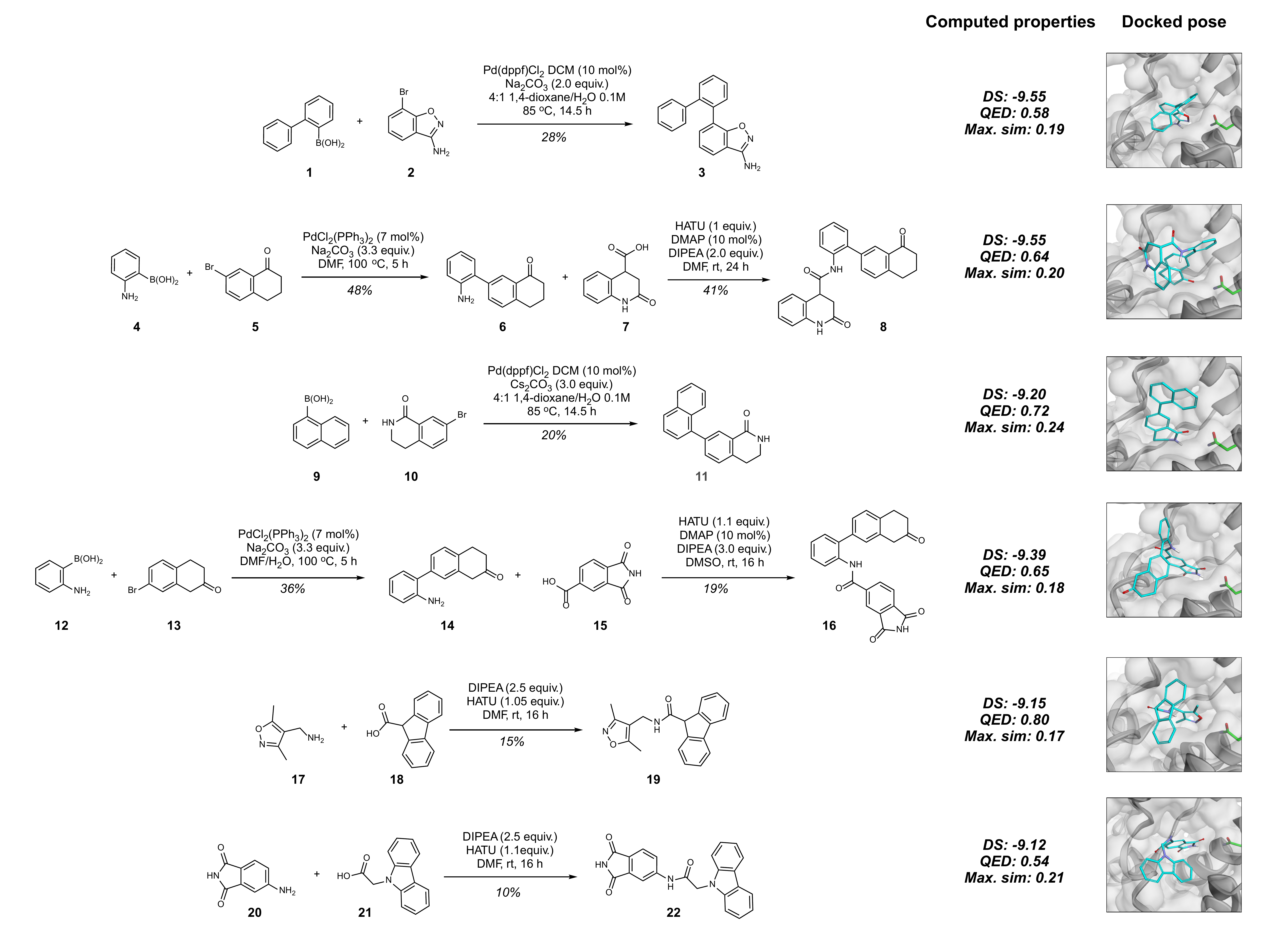}
    \caption{Synthesis of the six selected compounds in the BRD4 study. Computed oracle properties and GNINA docked poses are also shown.}
    \label{extended:synthesis}
\end{figure}

\begin{sidewaystable}[!ht]
\centering
\caption{\textbf{Enforcing \textit{all} reactions are from the specified set}. The enforced reaction(s) is present in all step(s) of the predicted route. The mean and standard deviation across 5 replicates are reported. \textbf{N} denotes the number of replicates (out of 5) that have at least one generated molecule satisfying all reaction constraints. The total number of molecules in each docking score interval pooled across all 5 replicates is denoted by \textbf{M} (only molecules satisfying all reaction constraints are considered). Within each of these docking score intervals, the \textbf{QED}, number of hydrogen-bond donors (\textbf{HBD}), ligand efficiency (\textbf{LE}), number of Bemis-Murcko scaffolds (\textbf{BMS}), internal diversity (\textbf{IntDiv1}), and number of circles (\textbf{\#Circles} threshold=0.75) are annotated. The oracle budget was fixed at 15,000.}
\label{table:enforce-all-metrics}
\renewcommand{\arraystretch}{1.5}
\setlength{\tabcolsep}{2.0pt}
\begin{tabular}{|l|c|c|c|c|c|c|c|}
\hline
\multirow{2}{*}{\textbf{Enforced}} & \multicolumn{3}{c|}{\textbf{Synthesizability}} & \multirow{2}{*}{\textbf{Rxn Steps}} & \multicolumn{3}{c|}{\textbf{Docking Score Intervals}} \\ 
 & \textbf{Non-synth} & \textbf{Synth} & \textbf{Synth} & & \textbf{DS $<$ -10} & \textbf{-10 $<$ DS $<$ -9} & \textbf{-9 $<$ DS $<$ -8} \\
 & & & \textbf{(constraints)} & & & & \\ \hline
& & & & & -10.35 $\pm$ 0.27 (M=1564) & -9.45 $\pm$ 0.26 (M=12230) & -8.56 $\pm$ 0.28 (M=19377) \\
& & & & & 0.89 $\pm$ 0.06 (\textbf{QED}) & 0.91 $\pm$ 0.05 & 0.90 $\pm$ 0.06 \\
& & & & & 1.11 $\pm$ 0.33 (\textbf{HBD}) & 1.09 $\pm$ 0.29 & 1.09 $\pm$ 0.30 \\
Amide & 1998 $\pm$ 276 & 13016 $\pm$ 277 & 9890 $\pm$ 467 & 1.21 $\pm$ 0.82 & 0.41 $\pm$ 0.02 (\textbf{LE}) & 0.39 $\pm$ 0.03 & 0.36 $\pm$ 0.03 \\
(N=5) & & & & & 1301 (\textbf{BMS}) & 8585 & 14849 \\
& & & & & 0.759 (\textbf{IntDiv1}) & 0.770 & 0.780 \\
& & & & & 15 (\textbf{\#Circles}) & 30 & 51 \\
\hline
& & & & & -10.35 $\pm$ 0.27 (M=2073) & -9.46 $\pm$ 0.27 (M=13042) & -8.57 $\pm$ 0.28 (M=18934) \\
& & & & & 0.88 $\pm$ 0.07 & 0.90 $\pm$ 0.06 & 0.90 $\pm$ 0.07 \\
& & & & & 1.09 $\pm$ 0.29  & 1.09 $\pm$ 0.30 & 1.10 $\pm$ 0.31 \\
Amide-Suzuki & 2008 $\pm$ 177 & 13000 $\pm$ 174 & 9790 $\pm$ 129 & 1.29 $\pm$ 0.85 & 0.40 $\pm$ 0.02  & 0.39 $\pm$ 0.03 & 0.36 $\pm$ 0.03 \\
(N=5) & & & & & 1918  & 9929 & 14340 \\
& & & & & 0.752 & 0.773 & 0.785 \\
& & & & & 21  & 29 & 59 \\
\hline
\end{tabular}%
\end{sidewaystable}

\begin{sidewaystable}[h]
\centering
\tiny
\caption{\textbf{Enforcing \textit{all} reactions are from the specified set and enforcing blocks}. The enforced reaction(s) is present in all step(s) of the predicted route. The mean and standard deviation across 5 replicates are reported. \textbf{N} denotes the number of replicates (out of 5) that have at least one generated molecule satisfying all reaction constraints. The total number of molecules in each docking score interval pooled across all 5 replicates is denoted by \textbf{M} (only molecules satisfying all reaction constraints are considered). Within each of these docking score intervals, the \textbf{QED}, number of hydrogen-bond donors (\textbf{HBD}), ligand efficiency (\textbf{LE}), number of Bemis-Murcko scaffolds (\textbf{BMS}), internal diversity (\textbf{IntDiv1}), and number of circles (\textbf{\#Circles} threshold=0.75) are annotated. The oracle budget was fixed at 15,000.}
\label{table:enforce-all-and-blocks-metrics}
\renewcommand{\arraystretch}{1.5}
\setlength{\tabcolsep}{2.0pt}
\begin{tabular}{|l|c|c|c|c|c|c|c|}
\hline
\multirow{2}{*}{\textbf{Reaction}} & \multicolumn{3}{c|}{\textbf{Synthesizability}} & \multirow{2}{*}{\textbf{Rxn Steps}} & \multicolumn{3}{c|}{\textbf{Docking Score Intervals}}\\ 
 & \textbf{Non-synth} & \textbf{Synth} & \textbf{Synth} & & \textbf{DS $<$ -10} & \textbf{-10 $<$ DS $<$ -9} & \textbf{-9 $<$ DS $<$ -8}\\
  & & & \textbf{(constraints)} & & & & \\ \hline
  
& & & & & -10.37 $\pm$ 0.26 (M=115) & -9.44 $\pm$ 0.28 (M=634) & -8.48 $\pm$ 0.27 (M=1643) \\
& & & & & 0.79 $\pm$ 0.06 \textbf{(QED)} & 0.81 $\pm$ 0.06 & 0.84 $\pm$ 0.06 \\
& & & & & 1.94 $\pm$ 0.24 \textbf{(HBD)} & 1.83 $\pm$ 0.38 & 1.31 $\pm$ 0.46 \\
Amide & 2920 $\pm$ 242 & 12087 $\pm$ 242 & 3947 $\pm$ 1156 & 1.45 $\pm$ 1.04 & 0.40 $\pm$ 0.02 \textbf{(LE)} & 0.37 $\pm$ 0.03 & 0.37 $\pm$ 0.04 \\
(N=5) & & & & & 113 \textbf{(BMS)} & 551 & 1379 \\
& & & & & 0.539 \textbf{(IntDiv1)} & 0.597 & 0.698 \\
& & & & & 1 \textbf{(Circles)} & 2 & 3 \\
\hline

& & & & & -10.20 $\pm$ 0.13 (M=7) & -9.36 $\pm$ 0.23 (M=450) & -8.46 $\pm$ 0.28 (M=2818) \\
& & & & & 0.80 $\pm$ 0.04 & 0.81 $\pm$ 0.06 & 0.84 $\pm$ 0.06 \\
& & & & & 1.57 $\pm$ 0.49 & 1.81 $\pm$ 0.40 & 1.64 $\pm$ 0.49 \\
Amide-Suzuki & 3144 $\pm$ 566 & 11863 $\pm$ 566 & 3069 $\pm$ 1691 & 1.53 $\pm$ 1.01 & 0.38 $\pm$ 0.03 & 0.34 $\pm$ 0.03 & 0.33 $\pm$ 0.03 \\
(N=5) & & & & & 7 & 429 & 2429 \\
& & & & & 0.570 & 0.649 & 0.719 \\
& & & & & 2 & 3 & 3 \\
\hline
\end{tabular}
\end{sidewaystable}

\clearpage

\begin{sidewaystable}[!ht]
\centering
\tiny
\caption{Docking MPO quantitative exact match rates in Freedom 4.0 given various degrees of synthesizability definition. The oracle budget was fixed at 15{,}000. \textbf{Non-synth} denotes no retrosynthesis route was returned for the molecule. \textbf{Synth} denotes a retrosynthesis route was returned but not necessarily satisfying the reaction constraints and \textbf{Synth Match} denotes the number of exact matches in this set. \textbf{Synth (constraints)} denotes a retrosynthesis route was returned \textit{and} satisfying the reaction constraints and \textbf{Synth (constraints) Match} denotes the number of exact matches in this set. \textbf{Exact Match Rate} denotes the percentage exact match in the \textbf{Synth (constraints)} set. Amongst the \textbf{Synth (constraints) Match} molecules, \textbf{Modes} is the number of molecules below the docking score thresholds that are also < 0.5 Tanimoto similarity to each other (radius=2, nBits=1024). \textbf{Yield} is the number of unique molecules below the docking score thresholds. All metrics are the mean and standard deviation across 5 replicates (0--4 inclusive). The \textbf{Direct Search Baseline} does not enforce reaction constraints and uses synthon-based search as a binary reward. The model is given 1.0 reward if RDKit synthon-based search verifies the generated molecule is in Freedom 4.0 and 0.0 otherwise.}
\vspace{1ex}
\label{tab:generative-virtual-screening}
\renewcommand{\arraystretch}{1.3}
\setlength{\tabcolsep}{3.5pt}
\begin{tabular}{|l|c|c|c|c|c|c|c|c|}
\hline
\textbf{Experiment} 
& \multicolumn{5}{c|}{\textbf{Synthesizability}} 
& \textbf{Exact Match} 
& \multicolumn{2}{c|}{\textbf{Modes (Yield) - QED}} \\
& \textbf{Non-synth} & \textbf{Synth} & \textbf{Synth Match} & \textbf{Synth} & \textbf{Synth} & \textbf{Rate} & \textbf{< -10 DS} & \textbf{< -9 DS} \\
& & & & \textbf{(constraints)} & \textbf{(constraints)} & & & \\
& & & & & \textbf{Match} & & & \\
\hline
\textbf{Direct Search} 
& N/A & N/A & N/A & N/A & 11123 $\pm$ 242\textsuperscript{a} 
& 0.74 $\pm$ 0.02 & 67 $\pm$ 15 & 255 $\pm$ 13 \\
\textbf{Baseline} 
& & & & & (15005 $\pm$ 4) & & (501 $\pm$ 280) & (3052 $\pm$ 769) \\
& & & & & & & 0.84 $\pm$ 0.01 & 0.85 $\pm$ 0.01 \\
\hline
\textbf{Baseline} 
& 5388 $\pm$ 171 & N/A & N/A & 9621 $\pm$ 174 & 3847 $\pm$ 712 
& 0.40 $\pm$ 0.08 & 25 $\pm$ 6 & 122 $\pm$ 19 \\
& & & & & & & (70 $\pm$ 31) & (715 $\pm$ 176) \\
& & & & & & & 0.83 $\pm$ 0.02 & 0.84 $\pm$ 0.02 \\
\hline
\textbf{All amide} 
& 5756 $\pm$ 208 & 9247 $\pm$ 207 & 8328 $\pm$ 209 & 8220 $\pm$ 202 & 7939 $\pm$ 244 
& 0.97 $\pm$ 0.01 & 15 $\pm$ 1 & 140 $\pm$ 18 \\
& & & & & & & (26 $\pm$ 5) & (980 $\pm$ 74) \\
& & & & & & & 0.83 $\pm$ 0.03 & 0.86 $\pm$ 0.01 \\
\hline
\textbf{All Suzuki} 
& 9053 $\pm$ 668 & 5950 $\pm$ 670 & 4155 $\pm$ 656 & 4452 $\pm$ 746 & 3389 $\pm$ 541 
& 0.78 $\pm$ 0.15 & 7 $\pm$ 3 & 47 $\pm$ 18 \\
& & & & & & & (16 $\pm$ 10) & (229 $\pm$ 158) \\
& & & & & & & 0.58 $\pm$ 0.01 & 0.64 $\pm$ 0.01 \\
\hline
\textbf{All Ester} 
& 8478 $\pm$ 424 & 6527 $\pm$ 425 & 175 $\pm$ 41 & 5079 $\pm$ 518 & 7 $\pm$ 3 
& 0.0015 $\pm$ 0.0007 & 0 $\pm$ 0 & 0 $\pm$ 0 \\
& & & & & & & (0 $\pm$ 0) & (0 $\pm$ 0) \\
& & & & & & & N/A & N/A \\
\hline
\textbf{Avoid amide} 
& 6179 $\pm$ 737 & 8826 $\pm$ 738 & 2630 $\pm$ 526 & 8686 $\pm$ 716 & 2553 $\pm$ 515 
& 0.29 $\pm$ 0.04 & 10 $\pm$ 4 & 69 $\pm$ 15 \\
& & & & & & & (17 $\pm$ 9) & (258 $\pm$ 117) \\
& & & & & & & 0.78 $\pm$ 0.06 & 0.81 $\pm$ 0.03 \\
\hline
\end{tabular}
\end{sidewaystable}


\clearpage
\includepdf[pages=-]{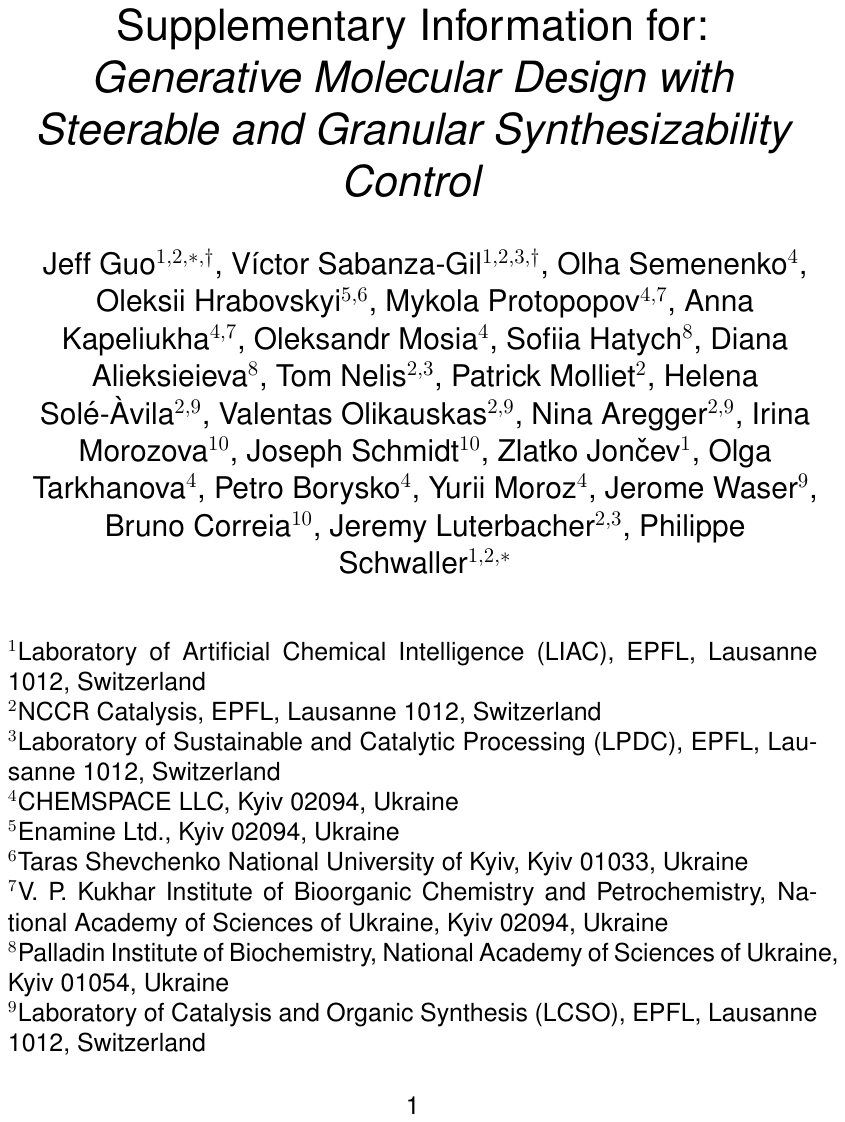}

\end{document}